\documentclass[twocolumn,floatfix,showpacs,showkeys,preprintnumbers,superscriptaddress,amsmath,amssymb]{revtex4}

\usepackage{graphicx}
\usepackage{dcolumn}% Align table columns on decimal point
\usepackage{bm}% bold math
\usepackage[latin1]{inputenc}
\usepackage[T1]{fontenc}
\usepackage{indentfirst}
\usepackage[]{amssymb}
\usepackage[]{amsmath}
\usepackage{subfigure}
\usepackage{lscape}
\usepackage{longtable}

\begin{document}

\title{Critical parameters for the partial coalescence of a droplet}

\author{T. Gilet}
\email{Tristan.Gilet@ulg.ac.be}
\affiliation{ GRASP, Physics Department B5, \\
University of Li\`ege, B-4000 Li\`ege, Belgium}

\author{K. Mulleners}
\affiliation{ GRASP, Physics Department B5, \\
University of Li\`ege, B-4000 Li\`ege, Belgium}

\author{J.P. Lecomte}
\affiliation{Dow Corning S.A.\\
Parc Industriel-Zone C \\
B-7180 Seneffe - Belgium}

\author{N. Vandewalle}
\author{S. Dorbolo}
\homepage{http://www.grasp.ulg.ac.be}
\affiliation{ GRASP, Physics Department B5, \\
University of Li\`ege, B-4000 Li\`ege, Belgium}

\date{\today}

\begin{abstract}
The partial coalescence of a droplet onto a planar liquid/liquid interface is investigated experimentally by
tuning the viscosities of both liquids. The problem mainly depends on four dimensionless parameters: the Bond
number (gravity vs. surface tension), the Ohnesorge numbers (viscosity in both fluids vs. surface tension), and
the density relative difference. The ratio between the daughter droplet size and the mother droplet size is
investigated as a function of these dimensionless numbers. Global quantities such as the available surface
energy of the droplet has been measured during the coalescence. The capillary waves propagation and damping are
studied in detail. The relation between these waves and the partial coalescence is discussed. Additional viscous
mechanisms are proposed in order to explain the asymmetric role played by both viscosities.
\end{abstract}


\maketitle

\section{Introduction}

Liquid droplets are more and more studied in the framework of microfluidic applications. Indeed, they allow to
manipulate and transport very small quantities of liquids. Droplets coalescence is probably the most convenient
way to mix liquids in microdevices without any power supply \cite{Wu:2004,Stone:2004}. Both control and
reproducibility of the droplet sizes are therefore very important. For instance, one of the main difficulties
encountered is to obtain a single small droplet, with a typical size lower than 100 $\rm{\mu m}$. Such a
challenge is also present when dealing with the petrol injection in car motors. The smaller the droplets are,
the more efficient the combustion is \cite{Thoroddsen:2006}. The partial coalescence is a suitable way to
achieve such a goal since it progressively empties a droplet \cite{Vandewalle:2006}.

\begin{figure}[h]
\includegraphics[width=0.85\columnwidth]{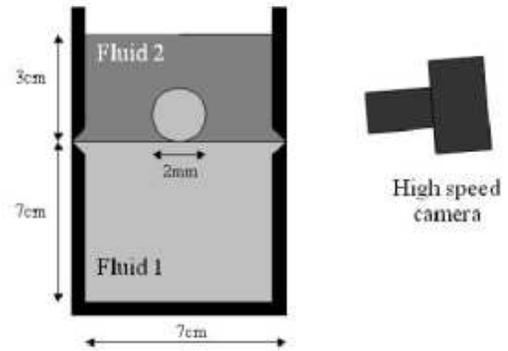}
\caption{\label{fig:SchExp} Experimental setup: A container is filled with two immiscible fluids. A droplet of
fluid 1 is falling through fluid 2, before coalescing with the fluid 1 bulk phase. The same experiment can be
conducted when fluid 1 is lighter than fluid 2 (the droplet is going up).}
\end{figure}

In order to investigate partial coalescence, Charles and Mason \cite{Charles:1960} have studied a system
composed by two immiscible fluids 1 and 2 of different densities (fluid 1 is heavier than fluid 2). This
experiment is schematically represented in Fig.\ref{fig:SchExp}. A thick layer of the lighter fluid 2 is placed
over the bulk of fluid 1. A millimetric droplet of fluid 1 is then dropped over the system. It crosses the fluid
2 layer and stays on the interface between fluid 1 and fluid 2 for a while depending on the viscosities.
Suddenly, the droplet coalesces. In given conditions that will be discussed, the droplet experiences a partial
coalescence. A smaller droplet remains above the 1,2 interface. The process may occur several times, like a
cascade: it is possible to get as much as six partial coalescences before the final total one
\cite{Vandewalle:2006}. Note that this experiment is totally equivalent to a droplet of fluid 1 (lighter than
fluid 2) coming up through fluid 2, when apparent gravity (gravity + buoyancy forces) is considered.

Charles and Mason attempted to explain the occurrence of partial coalescence by considering the ratio of
dynamical viscosities in both fluids. But this only parameter is not sufficient to fully understand partial
coalescence mechanisms. This experiment was studied in 1993 by Leblanc \cite{Leblanc:1993}. He was interested in
the stability of emulsions. Indeed, partial coalescence considerably slows down the gravity-driven phase
separation of two immiscible liquids, for instance when dealing with petrol demulsification. Leblanc identified
every important parameters for a partial coalescence prediction, but his work remained unpublished. In 2000,
Thoroddsen and Takehara \cite{Thoroddsen:2000} studied this phenomenon in more details for an air/water
interface. Same conclusions as Leblanc emerged concerning the impact of viscosity on the flow. In the same time,
partial coalescence with a surfactant addition was described by Pikhitsa and Tsargorodskaya
\cite{Pikhitsa:2000}. More recently, in 2006, Blanchette and Bigioni \cite{Blanchette:2006} have explained the
mechanism of pinch-off in partial coalescence by using a subtle combination of both experimental and numerical
techniques. The convergence of capillary waves on the top of the droplet seemed to have a crucial importance on
the coalescence outcome. Several other studies have been made about partial coalescence in 2006: Honey and
Kavehpour \cite{Honey:2006} have considered the bouncing height of the daughter droplet. Aryafar and Kavehpour
\cite{Aryafar:2006} have focused on the time scales of the partial coalescence. Chen, Mandre, and Feng
\cite{Chen1:2006} attempted to model these time scales, as well as the ratio between the daughter and the mother
droplets. They recently extended their study to polymeric liquids \cite{Chen2:2006}.

Little is known about microflows occurring during the coalescence process. First, traditional experimental
techniques are difficult to be applied at droplet scale. However, PIV experiments on droplet coalescence were
made by Mohamed-Kassim and Longmire \cite{Mohamed-Kassim:2004}. Unfortunately, the droplet was too large to
allow partial coalescence. The simplest way to get experimental information about microflows is to observe the
interface position, as in \cite{Thoroddsen:2005} for instance. On a numerical point of view, free surface flows
with changing topology (such as coalescence, break-down, and pinch-off) are really difficult and expensive to
compute. Despite these difficulties, a lot of information have already been collected about microflows in a
coalescence process.

The present paper consists in an experimental study made over many coalescence events, with various initial
droplet sizes, viscosities, and densities of both fluids. Two main aspects are investigated: the ratio between
the daughter and the mother droplets and the role of capillary waves on partial coalescence criteria. Moreover,
using image processing, global variables such as the available potential surface energy are recorded during the
coalescence process. Their evolution with time raise new questions and new challenges. The relation between
capillary waves and partial coalescence is discussed. In appendix, the number of successive partial coalescences
is theoretically computed for different pairs of fluids.

\section{Experimental setup}

Our experimental setup consists in a glass vessel which is opened at the top. This container is half filled with
the fluid 1. A centimetric layer of fluid 2 is then gently poured over the fluid 1 (see schematic view in
Fig.\ref{fig:SchExp}). A horizontal guide groove is dug in the vertical borders of the container in order to
reduce the meniscus at the oil/water interface (and in order to get a quasi-planar interface), as interface
curvature is known to modify the drainage time. Moreover, according to Blanchette \cite{Blanchette:2006}, it can
also influence the result of partial coalescence. A syringe is used to create a small droplet of fluid 1,
colored by methylene blue for visualization purpose. By changing the needle diameter, the initial droplet radius
$R_i$ can be changed. Between each experiment, the setup is washed with acetone to avoid interface
contamination.

The cascade of partial coalescences is studied using a set of different silicon oils for the fluid 2
(Trimethylsilyl terminated polydimethylsiloxane, under the tradename Dow Corning 200 fluid). Their kinematic
viscosity $\nu_2$ can be easily tuned from $\nu_2$=0.65 to 100 cSt. Mixtures are made in order to obtain
intermediate viscosities. The fluid 1 is made of a mixture of water and alcohol (glycerol or ethanol). The
kinematic viscosity $\nu_1$ may be modified by the alcool/water ratio. Densities are also changed (see Table
\ref{tab:PropFluids}). The interfacial tension is measured by the pendant drop method. It is approximately 45
mN/m for any oil/water interfaces (a 5mN/m error is considered for each error bar). It is roughly 40mN/m when
dealing with a glycerol/water mixture. The addition of ethanol greatly decreases the interfacial tension. It is
approximately 25mN/m for 10\%-ethanol, 19mN/m for 20\%-ethanol, 13mN/m for 30\%-ethanol, and 9mN/m for
40\%-ethanol. We have explored the different regimes of partial and total coalescence.

\begingroup
\squeezetable
\begin{table}
\begin{ruledtabular}
\begin{tabular}{cdd}
Liquid                  & \rho (kg/m^3) & \nu (cSt) \\
\hline
W                       & 1000          & 0.893     \\
\hline
87.5\%W + 12.5\%G       & 1030          & 1.16      \\
75\%W + 25\%G           & 1063          & 1.64      \\
62.5\%W + 37.5\%G       & 1093          & 2.65      \\
50\%W + 50\%G           & 1127          & 4.74      \\
25\%W + 75\%G           & 1195          & 30.1      \\
\hline
90\%W + 10\%E           & 983           & 1.35      \\
80\%W + 20\%E           & 969           & 1.82      \\
70\%W + 30\%E           & 954           & 2.23      \\
60\%W + 40\%E           & 934           & 2.47      \\
\hline \hline
DC-0.65cSt              & 760           & 0.65      \\
DC-1.5cSt               & 830           & 1.5       \\
DC-5cSt                 & 915           & 5         \\
DC-10cSt                & 934           & 10        \\
DC-50cSt                & 960           & 50        \\
DC-100cSt               & 965           & 100       \\
\hline
80\%DC-10cSt + 20\%DC-100cSt  & 940           & 19.5      \\
60\%DC-10cSt + 40\%DC-100cSt  & 946           & 30.4      \\
40\%DC-10cSt + 60\%DC-100cSt  & 953           & 46.4      \\
20\%DC-10cSt + 80\%DC-100cSt  & 959           & 69.7
\end{tabular}
\end{ruledtabular}
\caption{\label{tab:PropFluids} Experimental setup: Fluid properties (G=Glycerol, E=Ethanol, W=water, DC=Dow
Corning 200 oil).}
\end{table}
\endgroup

In order to ensure that no significant microflows are present at the beginning of the coalescence process, the
experiments for which the film drainage time is less than one second are rejected. The vessel is big enough to
avoid parasite reflections of capillary waves on the walls during the coalescence.

A fast video recorder (Redlake Motion Pro) is placed near the surface. A slight tilt (less than 10°) is needed
to see the bottom of the droplet, since the interface is curved by the weight of the droplet. Movies of the
coalescence have been recorded up to 2000 frames per second. Since the partial coalescence, scaled on the
capillary time, is about 10 ms, a whole coalescence process is usually captured by about 20 images. The pixel
size is about 30 $\rm{\mu m}$. In normal conditions, a millimetric droplet is represented by about 30 pixels.
About 150 coalescence events have been recorded. The droplet/camera distance is constant. Distances on the
picture are measured with a relative error less than 5\%. Since this error on distances is constant during a
single coalescence experiment, ratios of distances are characterized by an error significantly smaller than
forecasted.

Initial and final horizontal radii are measured for every experiment. In order to get information about the
partial coalescence process, the whole interface is tracked on each snapshot for each experiments. Of course, we
suppose that the interface (and even the whole flow) is axisymmetric. It is then possible to estimate global
measurements such as the volume above the interface or the surface potential energy. An original, highly robust,
and threshold-less interface detection method has been developed. Note that the reflected light can be detected
as a part of the interface, generating high discontinuities in interface shape. These discontinuities are
detected and deleted in the post-processing.

\section{Dimensional analysis} \label{section:DimAnal}

Once the thin film of fluid 2 below the droplet is broken, the dynamics of partial coalescence is
macroscopically governed by only three kinds of forces: interfacial tension, gravity, and viscosity forces in
both fluids. There are seven macroscopic parameters: surface tension $\sigma$, densities $\rho_1$ and $\rho_2$,
kinematic viscosities $\nu_1$ and $\nu_2$, gravity acceleration $g$, and the initial droplet radius $R_i
(i=1,2)$.

As Dooley et al. \cite{Dooley:1997} or Thoroddsen and Takehara \cite{Thoroddsen:2000} have shown, the partial
coalescence process is usually scaled by the capillary time. This time is the result of the balance between
interfacial tension and inertia. It is given by
\begin{equation}
\tau_{\sigma} = \sqrt{\frac{\rho_m R_i^3}{\sigma}},
\end{equation}
where $\rho_m$ is defined as the mean density $\frac{\rho_1 + \rho_2}{2}$. Therefore, the interfacial tension
has to be the main force for partial coalescence to occur. In other words, a self-similar process is only
possible when one force is dominant (the surface tension). That means that there is no natural length scale
related to the balance of two forces.

According to the $\pi-$theorem (Vaschy-Buckingham), it is possible to build only four independent dimensionless
numbers. Three of them are derived from the ratio of the characteristic time scales of the different forces with
the capillary time. The gravity time is given by
\begin{equation}
\tau_{g} = \sqrt{\frac{R_i}{g'}},
\end{equation}
($g'$ is the apparent gravity experienced by the droplet), and the viscosity times by
\begin{equation}
\tau_{\nu 1,2} = \frac{R_i^2}{\nu_{1,2}}.
\end{equation}

The Bond number is the square of the ratio between the capillary time and the gravity time
\begin{equation} \label{equation:Bo}
Bo=\frac{(\rho_1-\rho_2)g R_i^2}{\sigma}.
\end{equation}

The Ohnesorge numbers are the ratio between the capillary time and the viscous times in both fluids
\begin{equation} \label{equation:Oh}
Oh_{1,2}=\frac{\nu_{1,2}\sqrt{\rho_m}}{\sqrt{\sigma R_i}}.
\end{equation}

The last dimensionless number can be the relative difference of density
\begin{equation}
\Delta \rho=\frac{\rho_1-\rho_2}{\rho_1+\rho_2}.
\end{equation}

Ideally, any dimensionless quantity can be expressed as a function of these four parameters, especially the
ratio $\Psi$ between the daughter droplet radius and the mother droplet radius
\begin{equation}
\frac{R_f}{R_i} = \Psi \biggl( Bo, Oh_1, Oh_2, \Delta \rho \biggr).
\end{equation}

For large droplets, gravity is known to be as important as surface tension. Therefore, gravity significantly
accelerates the emptying of the mother droplet \cite{Leblanc:1993,Blanchette:2006}. Moreover, it flattens the
initial droplet. The ratio $\Psi$ is monotonically decreasing with $Bo$, and it suggests that a critical Bond
number $Bo_c(Oh_1,Oh_2,\Delta \rho)$ should exist for which $\Psi=0$ for $Bo>Bo_c$: coalescence becomes total.

As shown by Blanchette and Bigioni \cite{Blanchette:2006}, the partial coalescence process is mainly due to the
convergence of capillary waves at the top of the droplet. These waves are generated at the beginning of the
coalescence, by the receding interface below the droplet. The viscosity forces in both fluids, mainly present
for smallest droplets, damp these capillary waves and inhibit the partial coalescence
\cite{Leblanc:1993,Thoroddsen:2000,Blanchette:2006}. The ratio $\Psi$ also monotonically decreases with both
increasing Ohnesorge numbers. Critical Ohnesorges numbers $Oh_{1c}$ and $Oh_{2c}$ may be defined; beyond which
coalescence becomes total.

On the other hand, when the Bond (resp. Ohnesorge) number is much smaller than the critical Bond (resp.
Ohnesorge) number, gravity (resp. viscosity) can be considered as negligible. The $\Psi$ function does not
depend on $Bo$ (resp. $Oh$) anymore. For droplets with negligible Bond and Ohnesorge numbers, $\Psi$ becomes a
function $\Psi_0$ that only depends on $\Delta \rho$, the relative difference of inertial effects generated by
the interfacial tension in both fluids. In these conditions, the process is self-similar since the densities do
not change between two successive partial coalescences.

The previous dimensional analysis is correct when the droplet is at rest at the beginning of the coalescence.
The radius is the only key parameter to describe the droplet. Such an hypothesis is true when microflows due to
the previous droplet fall are damped out. The timescale of these flows can be roughly estimated as the time
needed for the thin film of fluid 2 to be drained out. When this time is much larger than the capillary time,
the initial microflows can be considered as negligible compared to the microflows generated by the coalescence
process. As already mentionned, this condition has been checked for each experiment of coalescence. Obviously,
the film drainage time is shorter for a droplet surrounded by a gaz than for a droplet surrounded by a liquid.
Therefore, it is easier to work with a liquid/liquid interface as evidenced by Mohammed-Kassim and Longmire
\cite{Mohamed-Kassim:2004}.

\section{Results}

\subsection{Qualitative description}

\begin{figure*}[htbp]
\subfigure[Partial coalescence: $Bo=0.15$, $Oh_1=0.002$, $Oh_2=0.013$]{\label{fig:SnapShotE30a}
\includegraphics[width=0.45\textwidth]{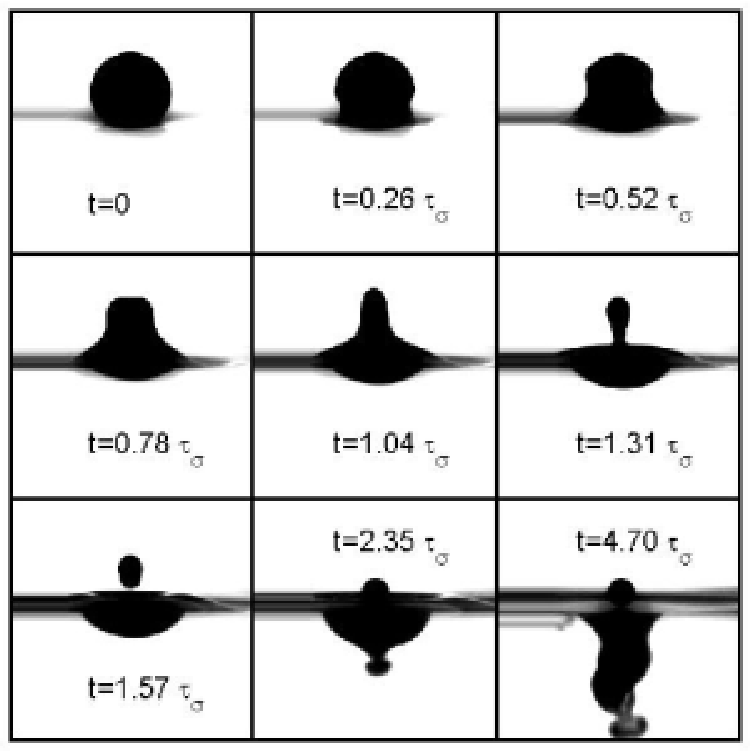}}
\subfigure[Total coalescence: $Bo=0.24$, $Oh_1=0.017$, $Oh_2=0.005$]{\label{fig:SnapShotE129a}
\includegraphics[width=0.45\textwidth]{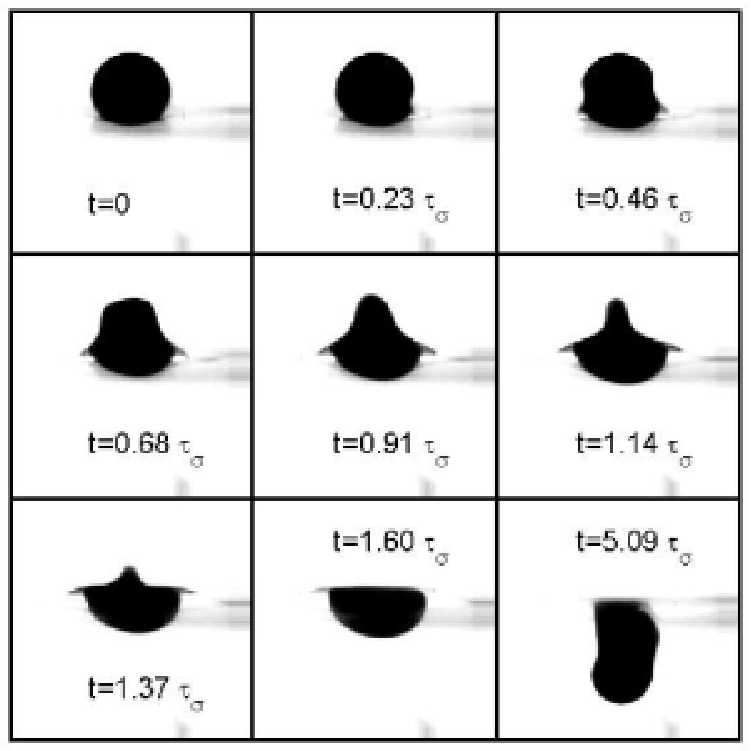}}
\subfigure[Partial coalescence: $Bo=0.03$, $Oh_1=0.003$, $Oh_2=0.16$]{\label{fig:SnapShotE54a}
\includegraphics[width=0.45\textwidth]{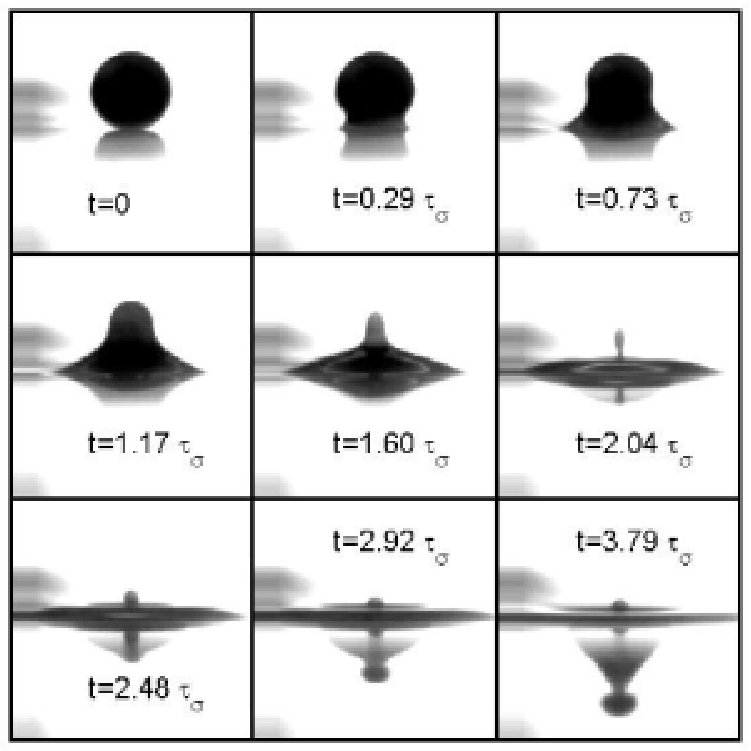}}
\subfigure[Total coalescence: $Bo=0.03$, $Oh_1=0.003$, $Oh_2=0.34$]{\label{fig:SnapShotE61a}
\includegraphics[width=0.45\textwidth]{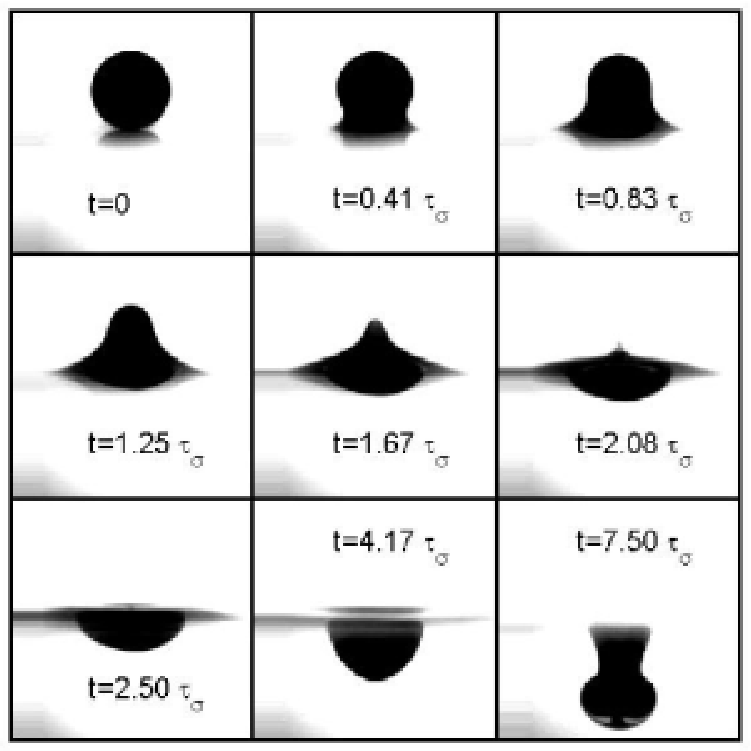}}
\caption{\label{fig:Snapshots} Partial and total coalescences of droplets at liquid/liquid interfaces. The whole
process is scaled by the capillary time $\tau_{\sigma}$. The first snapshot (Fig. \ref{fig:SnapShotE30a})
corresponds to a partial coalescence when both Ohnesorges are small. The second (Fig. \ref{fig:SnapShotE129a})
is a total coalescence mainly due to the high value of $Oh_1$. The both last snapshots, Fig.
\ref{fig:SnapShotE54a} and \ref{fig:SnapShotE61a}, depict a partial coalescence for an intermediate $Oh_2$, and
a total coalescence when the $Oh_2$ is high enough respectively.}
\end{figure*}

The different stages of a coalescence are shown on the snapshots of Fig. \ref{fig:Snapshots}. Four experiments
are presented. They have been chosen to evidence the influence of Ohnesorge numbers on the coalescence.

The first experiment (Fig. \ref{fig:SnapShotE30a}) corresponds to a quasi self-similar partial coalescence,
where the Bond number has a slight influence on the result, while the Ohnesorge numbers are negligible.

The film rupture usually occurs where the film is the thinnest. This latter is off-center, as it was shown
experimentally by Hartland \cite{Hartland:1967} and theoretically by Jones and Wilson \cite{Jones:1978}. This
off-centering effect is mainly due to the presence of a surrounding fluid, which generates an overpressure at
the center while it is drained outward.

When the film is broken, it quickly opens due to high pressure gradients created by surface tension near the
hole. Thoroddsen et al. \cite{Thoroddsen:2005} have mentioned that for water droplets surrounded by air, the
hole opening is dominated by inertia instead of viscosity. In this case, Eggers et al. \cite{Eggers:1999} have
proposed the following scaling law for the hole radius
\begin{equation} \label{equation:Eggers2}
r_h \sim \biggl( \frac{\sigma R_i}{\rho_1} \biggr)^{1/4} t^{1/2}.
\end{equation}
This law is based on a balance between inertia and surface tension; the biggest curvature is given by
$R_i/r_h^2$. It is possible to rewrite this equation with the capillary time
\begin{equation}
\biggl( \frac{r_h}{R_i} \biggr)^2 \sim \frac{t}{\tau_{\sigma}}.
\end{equation}
Measurements of Menchaca et al. \cite{Menchaca:2001}, Wu et al. \cite{Wu:2004}, and Thoroddsen et al.
\cite{Thoroddsen:2005} confirm this scaling. When the surrounding fluid is viscous oil instead of air, one can
expect that viscosity effects are as much important as inertia effects, and the mentioned scaling can become
obsolete.

During the retraction, a part of the bulk phase of fluid 1 comes up into the droplet, as shown by Brown and
Hanson \cite{Hanson:1967}. They observed a slightly colored daughter droplet when using a colorless mother
droplet on a colored bath. This upward movement is depicted in \cite{Dooley:1997}.

Next, the emptying droplet takes a column shape, and then experiences a pinch-off that leads to the formation of
the daughter droplet. Charles and Mason \cite{Charles:1960} have suggested this pinch-off was due to a
Rayleigh-Plateau instability. Recently, Blanchette and Bigioni \cite{Blanchette:2006} have shown that this
hypothesis was wrong. They have solved Navier-Stokes equations for a coalescing water droplet surrounded by air.
When the top of the droplet reaches its maximum height, they stop the numerical simulation, set velocities and
pressure perturbations to zero, and restart the computation. The coalescence, normally partial, was observed to
be total with this flow reset. The pinch-off cannot be due to a Rayleigh-Plateau instability.

Leblanc \cite{Leblanc:1993}, Thoroddsen et al. \cite{Thoroddsen:2000}, and Mohamed-Kassim et al.
\cite{Mohamed-Kassim:2004} have already noted the existence of capillary waves generated at the bottom of the
droplet after the film break-down. A part of these waves propagates far away, on the planar interface. The other
part climbs over the droplet and converges at the top. According to Blanchette \cite{Blanchette:2006}, such a
convergence greatly deforms the droplet and delays its coalescence, the horizontal collapse (the pinch-off) can
occur before the emptying. Gravity balances this delay by accelerating the emptying of the droplet. In other
words, too much gravity effects lead to a total coalescence. Viscosities of both fluids damp the capillary
waves, the convergence effect is reduced, and coalescence can also become total (Fig. \ref{fig:SnapShotE129a}
and \ref{fig:SnapShotE61a}). The precise role played by viscosities in the capillary waves damping will be
detailed in this paper.

The time at which the pinch-off occurs vary with the Bond and the Ohnesorge numbers. It is usually between 1.4
and 2 $\tau_{\sigma}$. After the pinch-off, the fluid below the droplet is ejected downward, due to the high
remaining pressure gradients. This creates a powerful vortex ring that go down through the fluid 1 (Fig.
\ref{fig:SnapShotE30a} and \ref{fig:SnapShotE54a}). Such a vortex is well described in \cite{Anilkumar:1991},
\cite{Shankar:1995} and \cite{Cresswell:1995}. Since this vortex is formed when the partial coalescence is
finished, it cannot have any influence on the coalescence outcome.

The shape evolution of the interface is relatively constant from one experiment to another. However, when $Oh_2$
is important (Fig. \ref{fig:SnapShotE54a} and \ref{fig:SnapShotE61a}), the interface takes a cusp-like shape
before diving away.

\subsection{The $\Psi$ law}

In this section, the ratio between the daughter droplet radius and the mother droplet radius (the $\Psi$
function) is investigated.

\subsubsection{Asymptotic regime}

Thanks to the dimensional analysis, we have shown that for negligible $Bo$, $Oh_1$, and $Oh_2$ (surface tension
is the only dominant force), $\Psi$ only depends on the density relative difference. Experimental results show
that one has
\begin{equation}
\Psi = \Psi_0 (\Delta \rho) \simeq cst. \simeq 0.45.
\end{equation}
In our experiments, the relative difference in density is always less than 20\%, and no relevant correlation
between $\Delta \rho$ and $\Psi$ has been underlined. Maybe the influence of the density relative difference is
higher for a liquid/gaz interface.

\subsubsection{Influence of the Bond number}

\begin{figure}[htbp]
\includegraphics[width=\columnwidth] {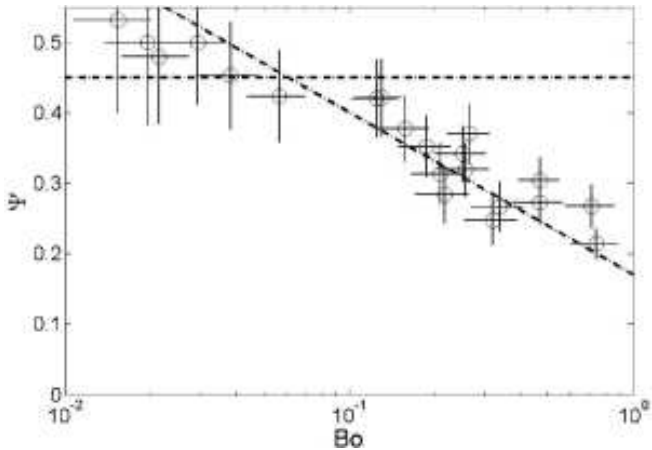}
\caption{\label{fig:PsiBo} Bond influence on the radius ratio $\Psi$. The Ohnesorge numbers are smaller than
$7.5 \times 10^{-3}$. Dashed lines are guides for the eyes.}
\end{figure}

Figure \ref{fig:PsiBo} shows the observed influence of the Bond number on $\Psi$, when the Ohnesorge numbers are
negligible (i.e. $<7.5 \times 10^{-3}$). For small Bond numbers, the asymptotic regime is obtained. As noted in
the dimensional analysis, an increase of the Bond number results in a decrease of $\Psi$. A cross over is
observed at $Bo \simeq 6 \times 10^{-2}$. According to our data, we cannot assess about the existence of any
critical Bond number $Bo_c$ for which $\Psi$ becomes zero. Mohamed-Kassim and Longmire
\cite{Mohamed-Kassim:2004} have reported total coalescences for Bond numbers equal to 10. However, both
Ohnesorge numbers were greater than 0.01, and viscosity played a significant part on the outcome of these
coalescences.

\subsubsection{Influence of the Ohnesorge numbers}

\begin{figure}[htbp]
\includegraphics[width=\columnwidth] {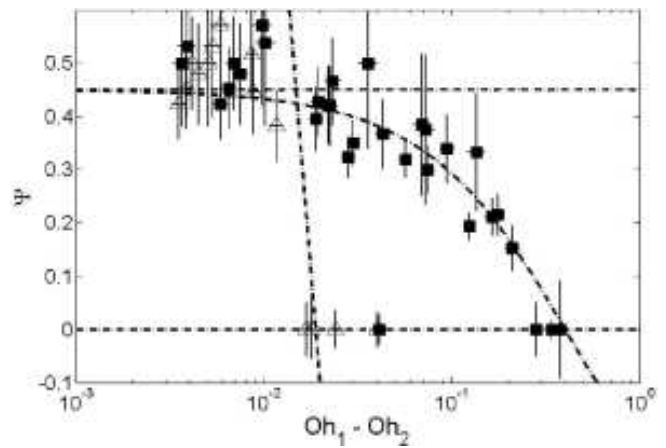}
\caption{\label{fig:PsiOh1Oh2} Ohnesorge 1 ($\vartriangle$) and Ohnesorge 2 ($\blacksquare$) influences on the
radius ratio $\Psi$. The Bond number is smaller than 0.1, and the other Ohnesorge number is smaller than $7.5
\times 10^{-3}$. Dashed lines are guides for the eyes.}
\end{figure}

In Fig.\ref{fig:PsiOh1Oh2}, the dependence of $\Psi$ on Ohnesorge numbers is plotted: Triangles correspond to
$Oh_1$, and black squares to $Oh_2$. For the $Oh_1$ (resp. $Oh_2$) curve, the selected data points correspond to
$Bo<0.1$ and $Oh_2<7.5\times 10^{-3}$ (resp. $Oh_1<7.5\times 10^{-3}$). Viscous dissipation in fluid 2 leads to
a smooth decreasing of $\Psi$ with $Oh_2$, starting from the asymptotic regime $\Psi \simeq 0.45$. The critical
Ohnesorge for partial coalescence $Oh_{2c}$ is about $0.3 \pm 0.05$. Leblanc \cite{Leblanc:1993} have observed
$Oh_{2c} \simeq 0.32$, which is in accordance with our results. The behavior of $\Psi$ with increasing $Oh_1$ is
totally different: the decrease of the ratio $\Psi$ is very sharp. It immediately vanishes to zero when $Oh_1 >
Oh_{1c} \simeq 0.02 \pm 0.005$. This critical value was already obtained by Blanchette \cite{Blanchette:2006} (a
$\sqrt{2}$ factor is needed to have the same definition of the Ohnesorge number), and thirteen years ago by
Leblanc \cite{Leblanc:1993} (who found 0.024). In the capillary waves scenario \cite{Blanchette:2006}, the roles
of both viscosities should be the same. The difference between their influences is then amazing and will be
discussed later.

\subsubsection{Conjugated influences}

In the previous sections, the $\Psi$ law has been investigated by varying a single parameter, both others being
negligible. Here, two parameters over three are varied, and the third one is taken negligible.

\begin{figure}[htbp]
\includegraphics[width=\columnwidth] {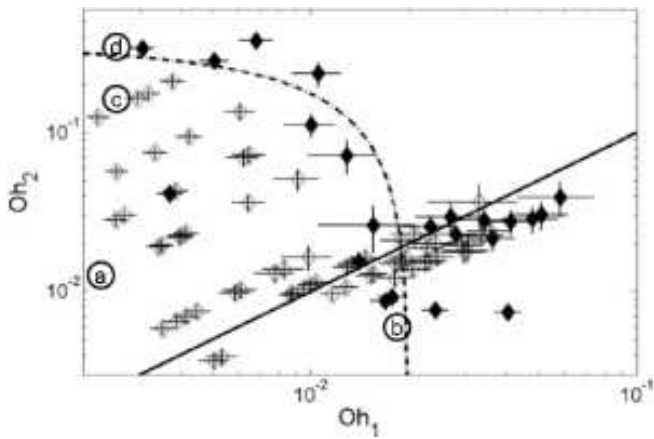}
\caption{\label{fig:Oh1Oh2} Phase plane $Oh_1$ - $Oh_2$ ($Bo < 0.1$). Experimental results: Total coalescence
($\blacklozenge$) vs. partial coalescence ($\lozenge$). The dashed line corresponds to the linear relationship
(\ref{equation:LinOh}). The solid line corresponds to the equality of both Ohnesorge numbers. Circled letters
indicate the position of the four snapshots (Fig. \ref{fig:Snapshots}).}
\end{figure}

The conjugated influence of viscosities is shown in Fig. \ref{fig:Oh1Oh2}. While the boundary between partial
and total coalescences is unambiguous for large and intermediate $Oh_2$, it is quite fuzzy for large $Oh_1$. It
seems that, considering for example $Oh_1=0.03$, the coalescence is total for small $Oh_2$ (as seen in the
previous section). Then, it can be partial for $Oh_2$ around 0.02, and it becomes total again for greater values
of $Oh_2$. This unexpected outgrowth of the partial coalescence regime roughly corresponds to the straight line
$Oh_1=Oh_2$.

In order to take both viscosities into account, Leblanc \cite{Leblanc:1993} suggested that the critical
parameter for partial coalescence occurrence is a linear combination of both Ohnesorge numbers. Partial
coalescence takes place when
\begin{equation} \label{equation:LinOh}
Oh_1 + 0.057 Oh_2 < 0.02.
\end{equation}
This law is compared to experimental results in Fig. \ref{fig:Oh1Oh2}. The law captures the general trend of the
partial-to-total transition, but fails in describing the large $Oh_1$ part of the diagram.

\begin{figure*}[htbp]
\subfigure[$Bo$ - $Oh_1$ ($Oh_2<7.5 \times 10^{-3}$)]{\label{fig:BoOh1}
\includegraphics[width=\columnwidth]{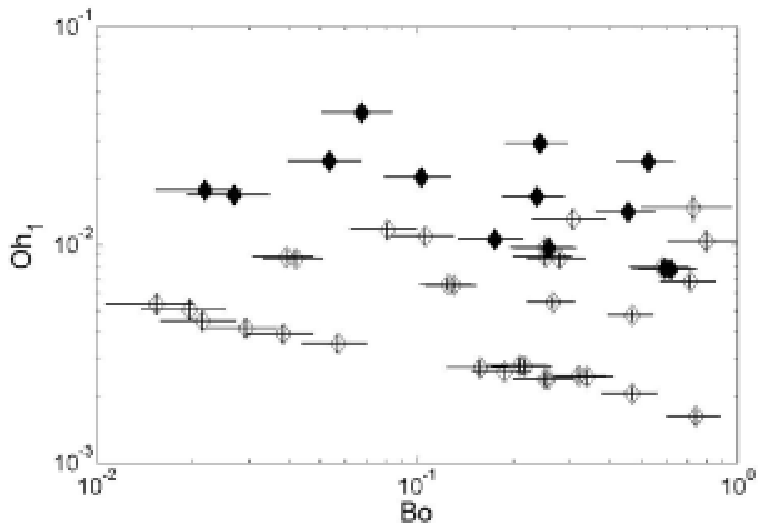}}
\subfigure[$Bo$ - $Oh_2$ ($Oh_1<7.5 \times 10^{-3}$)]{\label{fig:BoOh2}
\includegraphics[width=\columnwidth]{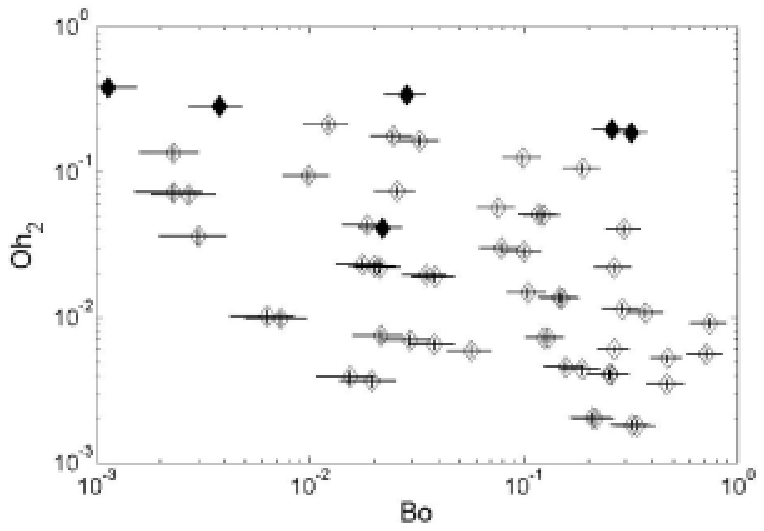}}
\caption{\label{fig:BoOh} Phase plane $Bo$ - $Oh_i$. Experimental results: Total coalescence ($\blacklozenge$)
vs. partial coalescence ($\lozenge$).}
\end{figure*}

Figure \ref{fig:BoOh1} and \ref{fig:BoOh2} indicate that the critical Ohnesorge numbers are slowly decreasing
when the Bond number becomes significant. This result agrees with data from Blanchette \cite{Blanchette:2006}.

\subsection{The global variables evolution \label{sec:Global}}

In order to get more information about the partial coalescence process, we have measured global quantities
associated to the detected interface on each experimental image: the volume $V$ of fluid 1 above the mean
interface level and the available surface potential energy $SPE$ (as in \cite{Thoroddsen:2005}).

\begin{figure*}[htbp]
\subfigure[Available surface energy]{\label{fig:SPE} \includegraphics[width=\columnwidth]{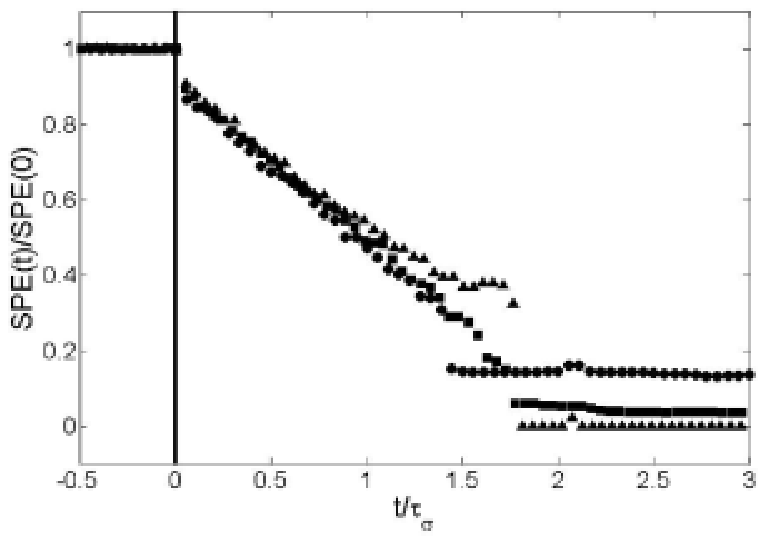}}
\subfigure[Volume above the mean interface level]{\label{fig:Vol} \includegraphics[width=\columnwidth]{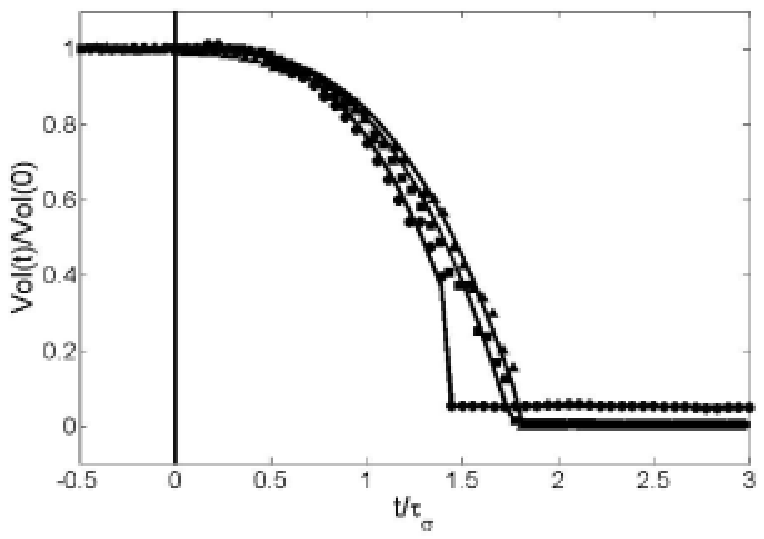}}
\caption{\label{fig:GlobEvol} Evolution of the global quantities for a self-similar partial coalescence
($\bullet$ - $Oh_2 \ll Oh_{2c}$), an attenuated partial coalescence ($\blacksquare$ - $Oh_2 \lesssim Oh_{2c}$),
and a total coalescence ($\blacktriangle$ - $Oh_2 > Oh_{2c}$).}
\end{figure*}

The available surface energy (resp. the volume $V$ of the droplet above the mean interface level), normalized by
its initial value, is plotted as a function of time in Fig.\ref{fig:SPE} (resp. in Fig.\ref{fig:Vol}). Results
of three experiments with different surrounding viscosities are presented: a partial coalescence ($Oh_2 \ll
Oh_{2c}$), an attenuated partial coalescence ($Oh_2 \lesssim Oh_{2c}$), and a total coalescence ($Oh_2 >
Oh_{2c}$). The Bond and the Ohnesorge 1 are negligible for the three experiments.

Figure \ref{fig:SPE} and \ref{fig:Vol} show that the volume and the surface energy decreases are roughly
independent of the coalescence issue.

During the main part of the coalescence process, the surface decrease is remarkably linear. Therefore, we can
assess that surface potential energy is converted into kinetic energy at a constant rate \footnote{The
gravitational energy is negligible since the Bond number is smaller than unity.} This rate does not seem to
depend on $Oh_2$, and corresponds to a 40\%-decrease in surface energy during the capillary time. Such a linear
decrease in surface energy was already observed when dealing with the initial hole opening, as seen previously.

The droplet starts to significantly empty at approximately $t=0.5 \tau_{\sigma}$. As seen in Fig. \ref{fig:Vol},
it is possible to correctly fit data by the following empirical law
\begin{equation} \label{equation:EvolVol}
\frac{V(t)}{V(t=0)}=1- C_V \biggl( \frac{t}{\tau_{\sigma}} \biggr)^3.
\end{equation}
According to the fitting, the coefficient $C_V$ is approximately 0.2, and experiences a slight increase with
$Bo$. Indeed, in Fig.\ref{fig:Vol}, the Bond number is higher for the partial coalescence experiment than for
the total one. Therefore, gravity enhances the droplet emptying. The flow rate $Q$ through the mean interface
level can be deduced by differentiate Eq.(\ref{equation:EvolVol})
\begin{equation}
\frac{Q \tau_{\sigma}}{V(t=0)} = 3 C_V \biggl( \frac{t}{\tau_{\sigma}} \biggr)^2.
\end{equation}

\subsection{Capillary waves}

As shown by Blanchette \cite{Blanchette:2006}, capillary waves that travel along the droplet are thought to be
responsible for the delay in the vertical collapse. Since this delay allows the pinch-off process to occur on
time (i.e. before the complete emptying), capillary waves have an important influence on the $\Psi$ function.
This section attempts to answer which modes make the convergence possible, and how they are damped by viscosity.

\subsubsection{Wave propagation}

Considering axisymmetric perturbations of an ideal, irrotationnal, and incompressible droplet of radius $R_i$,
it is possible to find the dispersion relationship of potential capillary waves on a spherical interface
\footnote{A similar development is made in the Fluid Mechanics course of Landau and Lifchitz
\cite{Landau:1959}.}
\begin{equation} \label{equation:Disp}
(\omega_l \tau_{\sigma})^2 = \frac{l(l^2-1)(l+2)}{2l+1+\Delta \rho},
\end{equation}
where $\tau_{\sigma}$ is the capillary time $\sqrt{\rho_m R_i^3/\sigma}$ and $\omega_l$ is the pulsation of mode
$l=k \times R_i$.

The interface is given by
\begin{equation}
\frac{R(t,\theta)}{R_i} = 1 + \sum_{l=0}^{\infty} A_l P_l(\cos \theta) \cos \omega_l t,
\end{equation}
where $A_l$ is the perturbation amplitude component associated to $P_l$, the Legendre polynomial of degree $l$.

Assuming relatively small wavelengths \footnote{When $l\geq8$, the approximation of Eq.(\ref{equation:Disp}) by
Eq.(\ref{equation:Disp2}) leads to a relative error smaller than 10\%, no matter the relative difference in
density.}, we can assess that $l = k R_i \gg 1$ and the dispersion relationship becomes
\begin{equation} \label{equation:Disp2}
(\omega_l \tau_{\sigma})^2 = \frac{l^3}{2}.
\end{equation}
Since this relation is independent on $\Delta \rho$, a difference in density between both fluids does not
generate any significant delay in the wave propagation to the top.

\begin{figure*}[htbp]
\includegraphics[width=\textwidth] {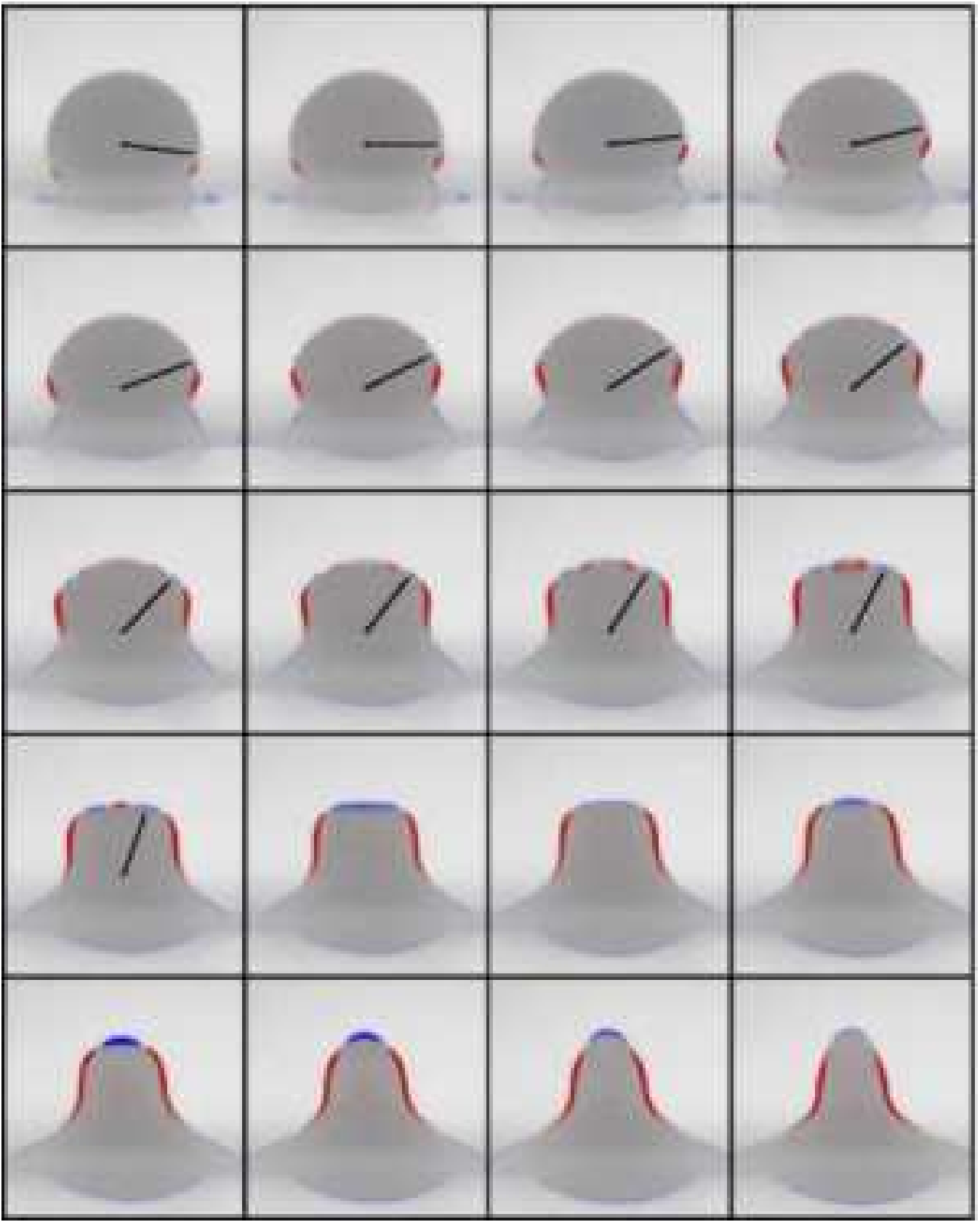}
\caption{\label{fig:OndesExpE30} (Color online) Capillary wave propagation obtained by image subtraction. The
blue interface is progressing toward fluid 2, while the red interface is receding. The time step is equal to
$0.045 \tau_{\sigma}$. $Bo=0.15$, $Oh_1=0.002$, $Oh_2=0.014$}
\end{figure*}

Experimentally, capillary waves can be detected by subtracting two successive images, as shown in Fig.
\ref{fig:OndesExpE30}. A progressing front appears in blue, while a receding one appears in red.

According to these pictures, the dominant mode can be estimated to $l \simeq 11 \pm 1$. It is created during the
initial hole opening, roughly 7.5° below the equator. The linear theory predicts a phase velocity
\begin{equation}
\frac{V_{\varphi} \tau_{\sigma}}{R_i} = \frac{\omega_l \tau_{\sigma}}{l}.
\end{equation}
Pictures give a constant angular phase velocity about 2.52 radians per capillary time unit, which corresponds to
$l \simeq 11$.

The propagation time $t_W$ of the waves from bottom to top is roughly proportional to $l^{-1/2}$. This means
that higher modes ($8<l$) arrive on top in the same time, while lower modes ($l<8$) arrive later, separately.
Therefore, the convergence cannot be due to these modes. Indeed, they are still on the way when the convergence
is observed.

\begin{figure}[htbp]
\includegraphics[width=\columnwidth] {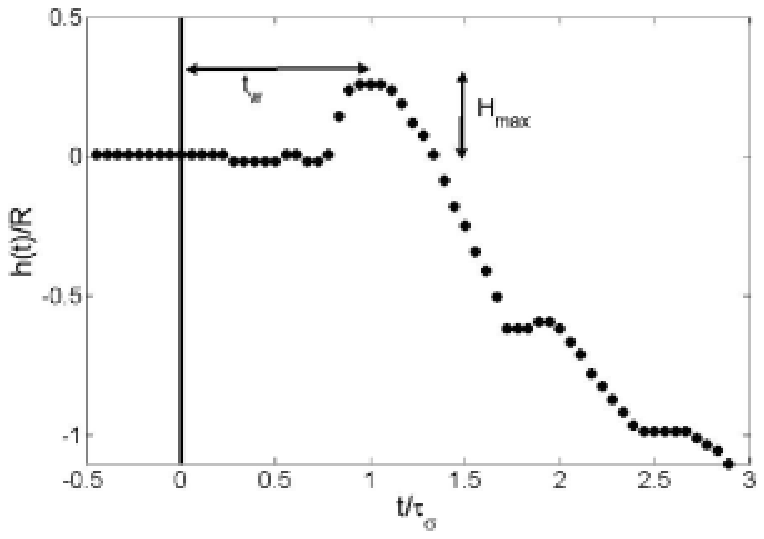} \caption{\label{fig:hsom} Evolution of the height of the droplet $h$ for a
self-similar partial coalescence ($\bullet$ - $Oh_2 \ll Oh_{2c}$), an attenuated partial coalescence
($\blacksquare$ - $Oh_2 \lesssim Oh_{2c}$), and a total coalescence ($\blacktriangle$ - $Oh_2 > Oh_{2c}$).}
\end{figure}

The relative elevation $h$ of the top of the droplet is plotted as a function of time in Fig. \ref{fig:hsom}.
Such a parameter was already studied by Mohamed-Kassim and Longmire \cite{Mohamed-Kassim:2004,Menchaca:2001}. At
the beginning of coalescence, the relative elevation is nearly zero since only the bottom of the droplet moves.
When the capillary waves are not too damped (partial coalescence), they converge at the top, and the elevation
increases until $H_{max}$, sometimes more than 30\% of the initial radius. The propagation time $t_W$ can be
estimated from these data as the time between the film rupture and the point of maximal elevation. Both
quantities $H_{max}$ and $t_W$ are illustrated in Fig. \ref{fig:hsom}.

\begin{figure}[htbp]
\includegraphics[width=\columnwidth] {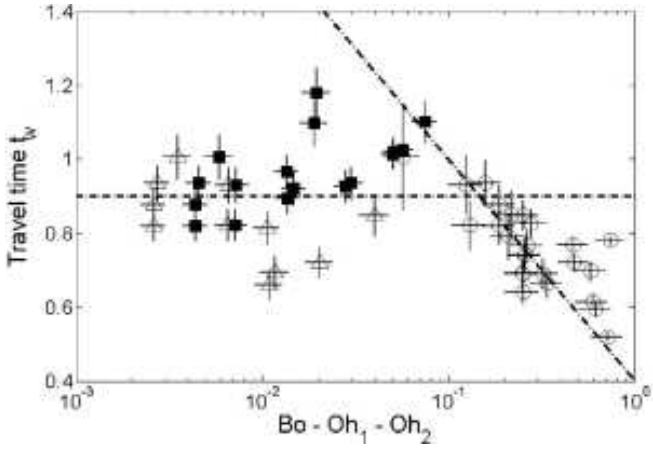}
\caption{\label{fig:CWTravelTime} Propagation time vs. $Oh_1 (\vartriangle)$, $Oh_2 (\blacksquare)$, and
$Bo(\circ)$. Dashed lines are guides for the eyes.}
\end{figure}

In Fig.\ref{fig:CWTravelTime}, it can be seen that this time is roughly $0.9 \tau_{\sigma}$, and it is
independent of $Oh_1$ and $Oh_2$.

On the other hand, a significant decrease is experienced with increasing Bond number. For example,
Mohamed-Kassim and Longmire \cite{Mohamed-Kassim:2004} obtained a propagation time approximately equal to 0.6
capillary time units. The dispersion relationship of surface-gravity waves on a planar interface between two
semi-infinite media is given by
\begin{equation}
(\omega \tau_{\sigma})^2 = \frac{l}{2} \biggl( l^2 + Bo \biggr),
\end{equation}
where $l$ is the angular momentum $k R_i$, $k$ the wave number, and $R_i$ the initial droplet radius. For
$Bo<1$, the influence of $Bo$ on this dispersion relationship is negligible compared to $l^2$. The phase
velocity is \textit{a priori} unaffected by the Bond number, and this cannot explain the observed decrease in
the propagation time. Another explanation could reside in the flattening of the droplet on the interface before
the coalescence when $Bo\lesssim Bo_c$ (especially since the horizontal radius is used for the computation of
the capillary time unit).

As a conclusion, convergence is mainly due to the effect of modes $l \geq 10$, the dominant one being $l=11$.
Waves converge to the top at approximately $0.9 \tau_{\sigma}$.

\subsubsection{Wave damping}

\begin{figure}[htbp]
\includegraphics[width=\columnwidth] {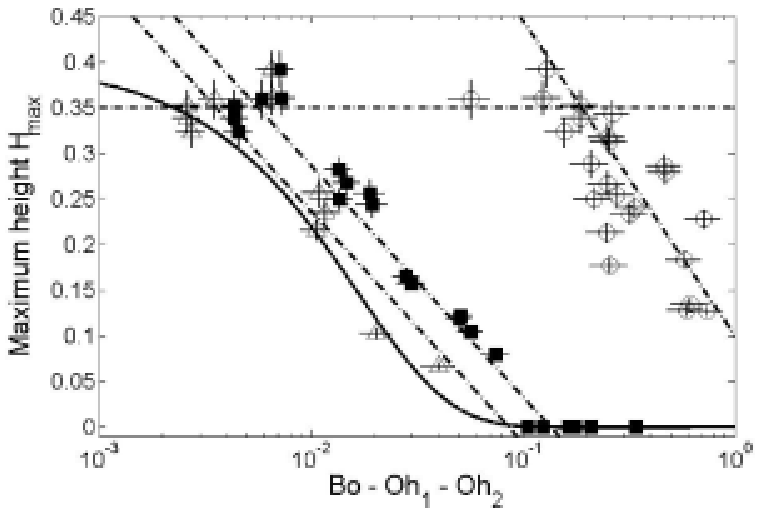}
\caption{\label{fig:Hmax} Maximum droplet elevation vs. Ohnesorge 1 $(\vartriangle)$, Ohnesorge 2
$(\blacksquare)$, and Bond $(\circ)$. Dashed lines are guides for the eyes. The solid line corresponds to the
forecast of the linear waves theory (Eq. (\ref{equation:HMax})).}
\end{figure}

Since the amplitude is small during the wave propagation, it is difficult to assess about wave damping by only
observing Fig. \ref{fig:OndesExpE30}. A better indicator is the dimensionless maximal elevation $H_{max}$ of the
top of the droplet. In Fig.\ref{fig:Hmax}, $H_{max}$ is plotted with respect to $Oh_1$, $Oh_2$ or $Bo$ (both
other parameters being negligible). It can be seen that both viscosities play the same part in damping the waves
\footnote{The 'damping' due to an increase of the Bond number is less relevant, since it is most probably the
result of a flattening of the droplet instead of a real dissipative mechanism.}: viscosity in fluid 1 is only
1.5 times more efficient than in fluid 2.

The viscous dissipation can be estimated according to the potential wave solution \cite{Landau:1959}
\footnote{Since the Reynolds number (the inverse of the Ohnesorge number) is high, the vorticity can be supposed
to be concentrated in a thin boundary layer. The potential solution is valid elsewhere and the viscous
dissipation can be estimated on the potential flow.}.

The velocity potential is given by
\begin{equation}
\Phi \simeq \left\{
\begin{array}{ll}
-A_l \frac{\omega_l}{l} r^l P_l(\cos \theta) \sin (\omega t) & \mbox{ in fluid 1} \\
A_l \frac{\omega_l}{l+1} r^{-(l+1)} P_l(\cos \theta) \sin (\omega t) & \mbox{ in fluid 2}.
\end{array}
\right.
\end{equation}
The local energy dissipation rate is analytically computed as
\begin{equation}
D = \frac{\partial v_i}{\partial x_j} \biggl( \frac{\partial v_i}{\partial x_j} + \frac{\partial v_j}{\partial
x_i} \biggr).
\end{equation}
The dimensionless total dissipated power is then given by
\begin{equation}
P = Oh_1 (1+\Delta \rho) \int_{V_1} D_1 dV + Oh_2 (1-\Delta \rho) \int_{V_2} D_2 dV.
\end{equation}
The total amount of mechanical energy is given by
\begin{equation}
E_m = (1+\Delta \rho) \int_{V_1} |v_1|^2 dV + (1-\Delta \rho) \int_{V_2} |v_2|^2 dV.
\end{equation}
According to Landau and Lifchitz \cite{Landau:1959}, the wave amplitude is decreasing as
\begin{equation}
A=A_0 e^{-\gamma_l t},
\end{equation}
where
\begin{equation}
\gamma_l = \frac{|P|}{2E_m}.
\end{equation}
On a spherical interface, the damping factor $\gamma_l$ is given by
\begin{equation}
\gamma_l = \frac{2 l +1}{2l+1+\Delta \rho} \biggl[ (l^2-1)Oh_1 (1+\Delta \rho) + l(l+2)Oh_2 (1-\Delta
\rho)\biggr].
\end{equation}

Therefore, according to the linear theory, one should expect
\begin{equation} \label{equation:HMax}
H_{max} = H_{max}^0 e^{-0.6 \gamma_l},
\end{equation}
where $H_{max}^0$ corresponds to the maximal height reached without viscous dissipation ($Oh_i \ll Oh_{ic}$). As
it can be seen on the snapshot in Fig. \ref{fig:OndesExpE30}, the 0.6 factor is the time needed by the waves to
travel from bottom to top (in capillary time units).

As shown in Fig. \ref{fig:Hmax}, the linear theory seems to give a representative picture of the wave damping.
However, it is not able to explain why fluid 2 is 1.5 times less efficient in damping waves than fluid 1.
Additional effects such as non-linearities and dissipation during the wave formation have to be taken into
account.

\subsubsection{Capillary waves and partial coalescence}

\begin{figure*}[htbp]
\includegraphics[width=\textwidth] {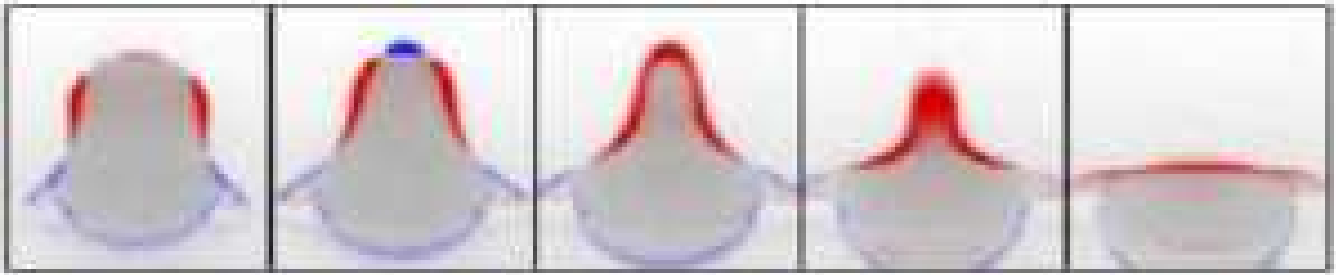}
\caption{\label{fig:OndesExpE130} (Color online) High viscosity in fluid 1 - Capillary waves and total
coalescence ($Bo=0.10$, $Oh_1=0.020$, $Oh_2=0.006$)}
\end{figure*}

\begin{figure*}[htbp]
\includegraphics[width=\textwidth] {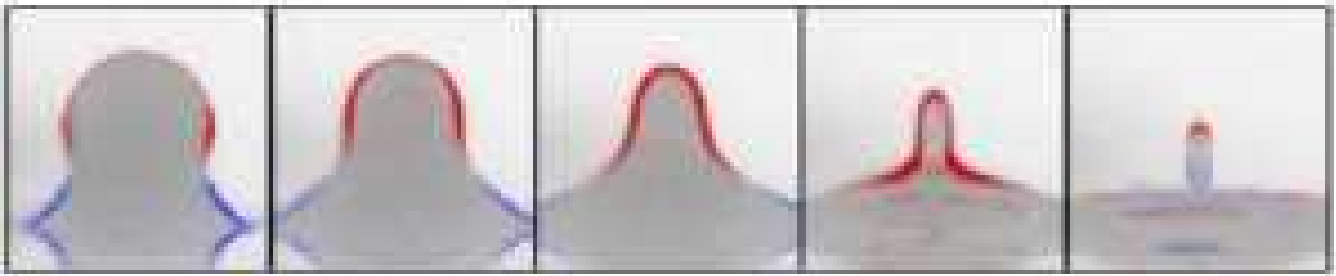}
\caption{\label{fig:OndesExpE54} (Color online) High viscosity in fluid 2 - No capillary waves and partial
coalescence ($Bo=0.03$, $Oh_1=0.003$, $Oh_2=0.16$)}
\end{figure*}

Although the delay in vertical collapse due to the convergence of capillary waves is important to get a partial
coalescence, it is not the only determinant factor. Indeed, in Fig. \ref{fig:OndesExpE130}, coalescence with an
important viscosity in fluid 1, the coalescence is total, despite the presence of capillary waves. On the other
side, in Fig. \ref{fig:OndesExpE54}, coalescence with an important viscosity in fluid 2, the coalescence is
partial, although the capillary waves are fully damped.

\section{Discussion}

The outcome of a coalescence is influenced by the viscosity of both fluids. Critical Ohnesorge numbers has been
identified for which a partial coalescence becomes total.

Until now, the partial coalescence was thought to be directly related to the convergence of capillary waves on
the top of the droplet. The critical Ohnesorge to get fully damped capillary waves is similar in fluid 1 and in
fluid 2, and is about $Oh_i \simeq 0.08$. Surprisingly, the critical Ohnesorge for the partial-to-total
coalescence transition is about 4 times higher in fluid 2 and 4 times lower in fluid 1: it is possible to
observe waves without partial coalescence and \textit{vice versa}.

Therefore, the convergence of capillary waves cannot be the only mechanism responsible for partial coalescence.
When $Oh_1$ is high, a mechanism has to be activated to enhance the emptying of the droplet, resulting in a
premature total coalescence. Inversely, when $Oh_2$ is high, another mechanism has to give advantage to the
horizontal collapse. However, when $Oh_2$ is too high, this advantage have to become a drawback.

When $Oh_1$ and $Oh_2$ are similar, the additional mechanisms balance themselves and the capillary waves remain
the only dominant factor. This could explain the outgrowth of the partial coalescence region on the solid line
of the $Oh_1-Oh_2$ map (Fig. \ref{fig:Oh1Oh2}).

\begin{figure}[htbp]
\includegraphics[width=\columnwidth] {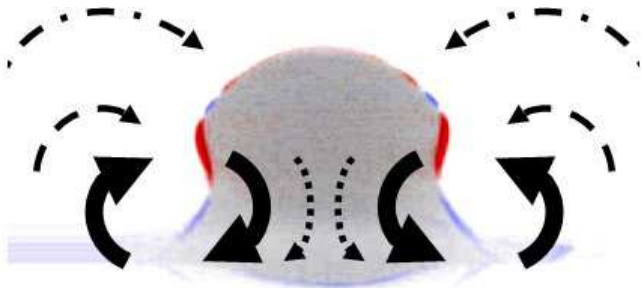}
\caption{\label{fig:Rotation} (Color online) Hypothetic schematic motion trends during the coalescence. The
solid arrows correspond to the nominal movement; the dotted arrows to the movement induced by a high viscosity
of fluid 1; the dashed arrows to the movement induced by an intermediate viscosity of fluid 2; and the dashdot
arrows to the movement induced by a high viscosity of fluid 2.}
\end{figure}

The main trend of movement during a coalescence is a powerful rotation, as it was shown in the PIV experiments
of Mohamed-Kassim et al. \cite{Mohamed-Kassim:2004}, as well in the numerical simulations of Blanchette et al.
\cite{Blanchette:2006} and Yue et al. \cite{Yue:2006}. This nominal movement is schematically represented with
solid arrows in Fig. \ref{fig:Rotation}.

The fluid motion is due to the conversion of interfacial energy into kinetic energy. As it was shown in section
\ref{sec:Global}, the kinetic energy supply rate is constant, and does not greatly depend on viscosities. This
energy is distributed in each fluid such as the normal velocity has to be continuous at the interface. When
viscosities are negligible, it gives rise to an 'ideal fluid' movement.

In presence of viscosity, the regions of high velocity gradients (visible on the numerical simulations of
Blanchette \cite{Blanchette:2006}) are diffusing momentum and energy. Since the motion is imposed by the
interface, the higher the viscosity is, the bigger the mass of fluid is displaced. In fluid 1, this additional
movement (represented by dotted arrows in Fig. \ref{fig:Rotation}) tends to accelerate the emptying of the
droplet. Note that other movements in fluid 1 can be enhanced by viscous diffusion when $Oh_1$ is increased,
especially the collapse of the final column as a whole.

In fluid 2, such a induced movement enhances the horizontal collapse. This could explain the amazingly thin
aspect of the column-shaped interface on the latest stages of coalescences when $Oh_2$ is high (see snapshots in
Fig. \ref{fig:SnapShotE54a} and \ref{fig:SnapShotE61a}). When $Oh_2$ is intermediate, the horizontal collapse is
confined to regions lower than the equator of the initial droplet, and this ensure a partial coalescence (dashed
arrows in Fig. \ref{fig:Rotation}, snapshot in Fig. \ref{fig:SnapShotE54a}). When $Oh_2$ increases, the viscous
diffusion entrains fluid at higher latitudes, and the vertical extension of the horizontal collapse is bigger:
coalescence becomes total (dashdot arrows in Fig. \ref{fig:Rotation}, snapshot in Fig. \ref{fig:SnapShotE61a}).
The boundary between these two opposite influences is approximately reached when energy is diffused on a scale
of the order of $R_i$, the initial radius of the droplet. The critical Ohnesorge in fluid 2 was found to be
about 0.32. This means that during the coalescence ($\simeq 1.5\tau_{\sigma}$), the energy is spread on a scale
roughly equal to $0.5 R_i$, which is coherent with the previous assumptions.

Further investigation is needed (for example a complete P.I.V. study) in order to confirm the existence of these
mechanisms, and to assess about the precise impact of viscosities on the flow.

\section{Conclusions}

Partial coalescence only depends on four dimensionless parameters, the Bond number (gravity/surface tension),
the Ohnesorge numbers in both fluids (viscosity/surface tension), and the density relative difference. The
partial coalescence can occur only when the Bond and the Ohnesorge numbers are below critical values $Bo_c$,
$Oh_{1c}$, and $Oh_{2c}$, since surface tension has to be the only dominant force. Otherwise, the coalescence is
total.

An experimental work on 150 coalescence events has been made in order to study the impact of viscosity and
gravity on the coalescence process. The droplets were filmed and the images post-processed in order to determine
the droplet radii, the capillary waves, but also quantities such as the potential surface energy along any
coalescence process. Surprisingly, it seems that this surface energy is converted into kinetic energy at a
constant rate, that does not depend on the coalescence outcome. This observation has no explanation until now.

The $\Psi$ law, which represents the ratio between the daughter and the mother droplet radii, has been
investigated as a function of $Bo$, $Oh_1$, and $Oh_2$. When these numbers are small enough, $\Psi$ becomes equal
to approximately 0.5. Gravity is known to accelerate the emptying of the droplet, while viscosity damps the
waves during their propagation. The critical Ohnesorge for the partial-to-total transition is more than one
order of magnitude higher in fluid 2 than in fluid 1.

Blanchette and Bigioni \cite{Blanchette:2006} have argued that partial coalescence was due to the waves
convergence on the top of the droplet. The propagation and the damping of these waves has been investigated. The
dominant modes are $10 \leq l \leq 12$. The damping of these waves is roughly the same in fluid 1 and in fluid
2. It cannot explain the huge difference between critical Ohnesorges in both fluids. Therefore, other viscous
mechanisms has been suspected to enhance or to avoid partial coalescence.

In appendix, a prediction of cascade features was made for a great number of different fluids. The record is
probably a 11-step cascade, when using an interface between mercury and another non-viscous liquid. Such a
cascade has not been observed for now. It would imply the use of advanced microscopy technique to observe the
late stages of the cascade.

Although partial coalescence begins to be understood, many questions remain unexplored. A P.I.V. study should
address a lot of these problems.

\begin{acknowledgments}
T.G. benefits of a FRIA grant (FNRS,Brussels); S.D. is a research associate of FNRS (Brussels). This work has
been financially supported by Colgate-Palmolive Inc. The authors would like to thank especially Dow Corning
S.A., Dr. G. Broze, Dr. H. Caps, Dr. M. Bawin, PhD Student S. Gabriel, and S. Becco for their fruitful help.
This work has been also supported by the contract ARC 02/07-293.
\end{acknowledgments}

\appendix

\section{Number and size of daughter droplets}

It is possible to estimate the maximum number of daughter droplets for a couple of fluids starting from their
physical properties (available in \cite{Kaye:1973} and \cite{Lide:2005} for instance). The maximum droplet size
(limited by gravity - Eq.(\ref{equation:Bo})) has a radius of the order of
\begin{equation}
R_M \simeq \sqrt{\frac{\sigma Bo_c}{2 \rho_m \Delta \rho g}},
\end{equation}
while the minimum droplet radius (viscosity limited - Eq.(\ref{equation:Oh})) is of the order of
\begin{equation}
R_{m} \simeq \max_{i=1,2} \biggl\{ \frac{\rho_m \nu_i^2}{\sigma Oh_{ci}^2} \biggr\}.
\end{equation}
Supposing that $\Psi \simeq 0.5$ during the whole cascade, the upper bound of the possibles numbers of steps in
the cascade $N$ is given by
\begin{equation}
\frac{\max(R_{m1},R_{m2})}{R_M} \simeq (0.5)^N.
\end{equation}
When $N=1$, the coalescence is total.

The capillary time can be estimated for the largest and the smallest droplet. Moreover, we have to ensure that
this time is always very short compared to the drainage time (when the droplet stays at rest on a thin film of
fluid 2). When the fluid 2 has a dynamical viscosity significantly smaller than the fluid 1, the film drainage
does not entrain the fluid 1 (neither in the droplet, nor in the bulk phase). The Reynolds'model
\cite{Leblanc:1993}, based on the lubrication equation, is of application. The lifetime $LT$ is given by
\begin{equation}
LT \approx \frac{3\mu_2 R_i^4}{4\pi g (\rho_1 - \rho_2) h_c^2},
\end{equation}
where $h_c$ is the critical film thickness when the film breaks. When dynamical viscosities are similar, the
film drainage entrains fluid 1, and the Reynolds'theory have to be replaced by Ivanov and Traykov's equation
\cite{Ivanov:1976}
\begin{equation}
LT \approx \biggl( \frac{\rho_1 \mu_1 R_i^2}{32 \pi^2 g^2 (\rho_1-\rho_2)^2 h_c^2} \biggr)^{1/3}.
\end{equation}
This equation is used to provide a crude estimation of the lifetime of the droplets in Table
\ref{tab:NumberStep}. Note that the drainage time is highly influenced by several factors: surfactant
\cite{Leblanc:1993}, electric fields \cite{Eow:2003}, vibrations \cite{Couder:2005}, and initial conditions
among others.

Tables \ref{tab:NumberStep} and \ref{tab:NumberStep2} give predicted orders of magnitude of $R_M$, $R_{m1}$,
$R_{m2}$, the associated lifetimes and capillary times, and the number of steps $N$ for different fluids 1. The
fluid 2 is air (A), water (W) or mercury (M). When fluid 2 is heavier than fluid 1, the droplet crosses the
interface from bottom to top. When the symbol * is used, fluids 1 and 2 are inverted (for example a droplet of
mercury surrounded by benzene instead of a droplet of benzene surrounded by mercury). The results in bold have
been approximately checked experimentally (by Charles and Mason \cite{Charles:1960}, Leblanc \cite{Leblanc:1993}
or our results).

% Tables 2 and 3 : Theoretical predictions
\begingroup
\squeezetable
\begin{turnpage}
\begin{table*}
\begin{ruledtabular}
\tiny{
\begin{tabular}{cccddddddddddd}
  Name  & Chemical  & Other & \rho_1      & \nu_1   & \sigma  & R_M     & LT_M    & \tau_{\sigma M} & R_{m1}  & R_{m2}  & LT_m    & \tau_{\sigma m} & N \\
        & formula   & phase & (kg/m^3)    & (cSt)   & (mN/m)  & (mm)    & (s)     & (ms)            & (\mu m) & (\mu m) & (ms)    & (\mu s)         & \\
  \hline
  \hline
  Acetic acid           & $C_2H_4O_2$   & A     & 1049  & 1.16 & 27.8   & 1.4   & 0.19  & 7.0   & 64    & 27    & 25    & 70    & 4 \\
  \hline
  Acetone (25°C)        & $C_3H_6O$     & A     & 787   & 0.39 & 23.5   & 1.5   & 0.14  & 7.2   & 6.3   & 25    & 9.4   & 17    & 5 \\
                        &               & M     &       &      & 390.0  & 1.5   & 0.022 & 7.7   & 6.9   & 1.4[-3] & 0.62 & 25    & 7 \\
                        &               & M*    &       &      &        &       & 0.097 &       & 0.56  & 0.017 & 0.50  & 0.056 & 11 \\
  \hline
  \textbf{Aniline}          & $C_6H_7N$     & A     & 1022  & 4.30 & 42.9   & 1.7   & 0.35  & 7.9   & 550   & 17    & 160   & 1.4[3] & 1 \\
                        &               & \textbf{W} &       &      & 6.1    & \textbf{4.4}   & 8.5   & 120   & \textbf{7.6[3]} & 1     & 12[3]  & 2.7[5] & \textbf{1} \\
                        &               & W*    &       &      &        &       & 5.1   &       & 410   & 19    & 1100  & 3400  & 3 \\

  \hline
  \textbf{Benzene}          & $C_6H_6$      & A     & 879   & 0.74 & 28.9   & 1.5   & 0.18  & 7.4   & 21    & 22    & 10    & 12    & 6 \\
                        &               & M     &       &      & 357    & 1.4   & 0.029 & 7.6   & 27    & 1.5[-3] & 21   & 20    & 5 \\
                        &               & M*    &       &      &        &       & 0.09  &       & 0.61  & 0.068 & 0.54  & 0.068 & 11 \\
                        &               & W     &       &      & 35.0   & \textbf{4.5} & 1.4   & 50    & 36    & 0.17  & 56    & 36    & 6 \\
                        &               & \textbf{W*} &       &      &        &       & 1.7   &       & \textbf{67}    & 0.091 & 100   & 90    & \textbf{6} \\
  \hline
  Carbon disulphide     & $CS_2$        & A     & 1263  & 0.30 & 32.3   & 1.4   & 0.12  & 7.0   & 4.3   & 28    & 9.2   & 21    & 5 \\
  \hline
  \textbf{Carbon tetrachloride} & $CCl_4$   & A     & 1594  & 0.61 & 27.0   & 1.1   & 0.12  & 6.4   & 10    & 39    & 12    & 41    & 4 \\
                        &               & \textbf{W} &       &      & 45.0   & \textbf{2.3} & 0.43  & 19    & 27    & 0.18  & 22    & 24    & 6 \\
                        &               & W*    &       &      &        &       & 0.37  &       & 72    & 0.067 & 37    & 100   & 5 \\
  \hline
  Chloroform            & $CHCl_3$      & A     & 1489  & 0.38 & 27.1   & 1.1   & 0.12  & 6.4   & 10    & 39    & 12    & 41    & 4 \\
                        &               & W     &       &      & 28.0   & 2.0   & 0.36  & 19    & 16    & 0.28  & 15    & 14    & 6 \\
                        &               & W*    &       &      &        &       & 0.38  &       & 110   & 0.041 & 56    & 250   & 4 \\
                        &               & M     &       &      & 357.0  & 1.5   & 0.034 & 8.0   & 7.7   & 1.6[-3] & 1.0  & 3.1   & 7 \\
                        &               & M*    &       &      &        &       & 0.099 &       & 0.64  & 0.019 & 0.57  & 0.074 & 11 \\
  \hline
  Cyclohexane (25°C)    & $C_6H_{12}$   & W     & 773   & 1.15 & 41.1   & 3.6   & 0.83  & 32    & 71    & 0.11  & 61    & 89    & 5 \\
                        &               & W*    &       &      &        &       & 0.91  &       & 43    & 0.18  & 47    & 41    & 6 \\
  \hline
  \textbf{DC200-0.65cSt (25°C)} &              & A     & 761   & 0.65 & 15.9   & 1.2   & 0.15  & 6.6   & 25    & 36    & 14    & 34    & 5 \\
                        &               & W     &       &      & \simeq 30 & \textbf{3} & 0.58 & \textbf{28} & 31 & 0.15 & 28   & 30    & 6 \\
                        &               & \textbf{W*} &     &      &        &       & \textbf{0.77} &   & \textbf{58} & 0.078 & 56  & 76    & \textbf{5} \\

  \hline
  \textbf{DC200-1.5cSt (25°C)} &               & A     & 830   & 1.50 & 17.6   & 1.2   & 0.20  & 6.6   & 130   & 36    & 45    & 240   & 3 \\
                        &               & W     &       &      & \simeq 30 & \textbf{3.5} & 1.1 & \textbf{37}   & 17    & 0.15  & 150   & 390   & 4 \\
                        &               & \textbf{W*} &     &      &        &       & \textbf{1.1} &       & \textbf{60}    & 0.43  & 72    & 82    & \textbf{5} \\
  \hline
  \textbf{DC200-10cSt (25°C)}  &               & A     & 934   & 10.0 & 20.1   & 1.2   & 0.37  & 6.7   & 5800  & 35    & 1000  & 6.8[4] & 1 \\
                        &               & W     &       &      & \simeq 30 & \textbf{5.7} & 6.0 & \textbf{77}  & 8100  & 0.16  & 7600  & 1.3[5] & 1 \\
                        &               & \textbf{W*} &       &      &        &       & \textbf{2.8}   &       & \textbf{64}    & 20    & 140   & 92    & \textbf{6} \\
  \hline
  \textbf{DC200-50cSt (25°C)}  &               & A     & 960   & 50.0 & 20.8   & 1.2   & 0.64  & 6.7   & 1.4[5] & 35    & 1.5[4] & 8.3[6] & 1 \\
                        &               & W     &       &      & \simeq 30 & \textbf{7.3} & 17 & \textbf{110}   & 2.0[5] & 0.16  & 1.6[5] & 1.7[7] & 1 \\
                        &               & \textbf{W*} &       &      &        &       & \textbf{4.6}   &       & 65    & \textbf{510}   & 780   & 2100  & \textbf{3} \\
\end{tabular}
}
\end{ruledtabular}
\caption{\label{tab:NumberStep}Theoretical prediction of cascades of partial coalescence events for different
couples of fluids. Fluid 1 is given in the first column while fluid 2 is air (A), water (W) or mercury (M). When
the symbol * is used, fluids 1 and 2 are inverted. Notation $a[b]$ corresponds to $a \times 10^b$.}
\end{table*}
\end{turnpage}
\endgroup

\begingroup
\squeezetable
\begin{turnpage}
\begin{table*}
\begin{ruledtabular}
\tiny{
\begin{tabular}{cccddddddddddd}
  Name  & Chemical  & Other & \rho_1      & \nu_1   & \sigma  & R_M     & LT_M    & \tau_{\sigma M} & R_{m1}  & R_{m2}  & LT_m    & \tau_{\sigma m} & N \\
        & formula   & phase & (kg/m^3)    & (cSt)   & (mN/m)  & (mm)    & (s)     & (ms)            & (\mu m) & (\mu m) & (ms)    & (\mu s)         & \\
  \hline
  \hline
  \textbf{DC200-100cSt (25°C)} &               & A     & 965   & 100  & 20.9   & 1.2   & 0.8   & 6.7   & 5.8[5] & 35    & 4.8[4] & 6.7[7] & 1 \\
                        &               & W     &       &      & \simeq 30 & \textbf{7.8} & 25 & \textbf{130}   & 8.2[5] & 0.16  & 5.5[5] & 1.3[8] & 1 \\
                        &               & \textbf{W*} &       &      &        &        & \textbf{5.3}  &       & 65    & \textbf{2000}  & 2200  & 1.7[4] & \textbf{1} \\
  \hline
  Di-ethyl ether        & $C_4H_{10}O$  & A     & 714   & 0.34 & 17.0   & 1.3   & 0.12  & 6.8   & 6.1   & 30    & 10    & 24    & 5 \\
                        &               & M     &       &      & 379.0  & 1.5   & 0.019 & 7.6   & 5.4   & 1.4[-3] & 0.47 & 1.7   & 8 \\
                        &               & M*    &       &      &        &       & 0.095 &       & 0.57  & 0.014 & 0.51  & 0.059 & 11 \\
                        &               & W     &       &      & 10.0   & 1.6   & 0.26  & 18    & 25    & 0.54  & 16    & 36    & 6 \\
                        &               & W*    &       &      &        &       & 0.47  &       & 210   & 0.062 & 120   & 920   & 2 \\
  \hline
  \textbf{Dodecane}         & $C_{12}H_{26}$ & \textbf{W} & 750 & 1.80 & 47.0   & \textbf{3.7}   & 0.9   & 30    & \textbf{150}   & 0.12  & 110   & 250   & \textbf{4} \\
                        &               & W*    &       &      &        &       & 0.89  &       & 47    & 0.38  & 49    & 43    & 6 \\
  \hline
  Ethanol               & $C_2H_6O$     & A     & 789   & 1.52 & 22.8   & 1.4   & 0.22  & 7.2   & 100   & 25    & 37    & 130   & 3 \\
  \hline
  Ethyl acetate         & $C_4H_8O_2$   & A     & 925   & 0.49 & 23.6   & 1.4   & 0.14  & 6.9   & 12    & 28    & 11    & 21    & 5 \\
  \hline
  Formic acid           & $CH_2O_2$     & A     & 1220  & 1.32 & 37.13  & 1.5   & 0.21  & 7.3   & 71    & 23    & 28    & 77    & 4 \\
  \hline
  \textbf{Glycerol}         & $C_3H_8O_3$   & A     & 1261  & 1185 & 63.4   & 1.9   & 2.4   & 8.2   & 3.5[7] & 14    & 1.7[6] & 2.1[10] & 1 \\
                        &               & M     &       &      & 370.0  & \textbf{1.5} & 0.45  & 7.9   & 7.0[7] & 1.5[-3] & 5.9[5] & 8.3[10] & 1 \\
                        &              & \textbf{M*} &       &      &        &       & 0.099 &       & 0.61  & 1.8[5] & 2400  & 1.0[7] & \textbf{1} \\
  \hline
  Mercury               & $Hg$          & A     & 13546 & 0.11 & 486.1  & 1.6   & 0.099 & 7.6   & 0.46  & 20    & 5.3   & 10    & 6 \\
            (25°C)      &               &       & 13534 & 0.11 & 471    & 1.6   & 0.098 & 7.5   & 0.46  & 22    & 5.6   & 12    & 6 \\
  \hline
  Methanol              & $CH_4O$       & A     & 791   & 0.75 & 22.6   & 1.4   & 0.17  & 7.2   & 25    & 25    & 12    & 17    & 5 \\
  \hline
  \textbf{n-Heptane}        & $C_7H_{16}$   & M     & 684   & 0.72 & 379.0  & 1.5   & 0.024 & 7.6   & 25    & 1.4[-3] & 1.6  & 17    & 5 \\
                        &               & M*    &       &      &        &       & 0.095 &       & 0.57  & 0.062 & 0.51  & 0.059 & 11 \\
                        &               & W     &       &      & 51.0   & \textbf{3.4} & 0.51  & 25    & 22    & 0.10  & 17    & 13    & 7 \\
                        &               & \textbf{W*} &       &      &        &       & 0.73  &       & \textbf{41}  & 0.054 & 38    & 34    & \textbf{6} \\
  \hline
  n-Hexane              & $C_6H_{14}$   & A     & 659   & 0.58 & 18.4   & 1.4   & 0.16  & 7.1   & 15    & 26    & 11    & 17    & 5 \\
  \hline
  n-Octane              & $C_8H_{18}$   & A     & 703   & 0.78 & 21.8   & 1.5   & 0.18  & 7.3   & 24    & 23    & 12    & 15    & 5 \\
                        &               & W     &       &      & 51.0   & 3.5   & 0.56  & 27    & 25    & 0.10  & 21    & 16    & 7 \\
                        &               & W*    &       &      &        &       & 0.77  &       & 42    & 0.063 & 40    & 35    & 6 \\
  \hline
  \textbf{Toluene}          & $C_7H_8$      & A     & 867   & 0.67 & 28.4   & 1.5   & 0.17  & 7.4   & 17    & 22    & 10    & 13    & 6 \\
                        &               & W     &       &      & 28.0   & \textbf{3.9} & 1.1   & 44    & 38    & 0.21  & 52    & 43    & 6 \\
                        &               & W*    &       &      &        &       & 1.4   &       & \textbf{83}  & 0.095 & 110   & 140   & \textbf{5} \\
  \hline
  Water                 & $H_2O$        & A     & 1000  & 1.00 & 72.8   & 2.3   & 0.26  & 9.0   & 17    & 9.8   & 10    & 6.0   & 7 \\
                        &               & A*    &       &      &        &       & 7.2[-3] &      & 3900  & 0.043 & 10    & 2.0[4] & 1 \\
         (25°C)         &               &       &       & 0.89 & 72.0   & 2.3   & 0.25  & 9.0   & 14    & 11    & 8.3   & 4.3   & 7 \\
\end{tabular}
}
\end{ruledtabular}
\caption{\label{tab:NumberStep2} Theoretical prediction of cascades of partial coalescence events for different
couples of fluids. Fluid 1 is given in the first column while fluid 2 is air (A), water (W) or mercury (M). When
the symbol * is used, fluids 1 and 2 are inverted. Notation $a[b]$ corresponds to $a \times 10^b$.}
\end{table*}
\end{turnpage}
\endgroup

We can see in Tables \ref{tab:NumberStep} and \ref{tab:NumberStep2} that, contrary to Thoroddsen's predictions
\cite{Thoroddsen:2006}, the coalescence of a mercury droplet surrounded by air is done in only 6 stages. The
maximum number of cascade steps is obtained for a mercury droplet surrounded by a non-viscous liquid, such as
acetone, benzene, heptane, chloroform or ether. No more than 11 stages will be observed in this case. The record
could be broken by working with giant droplets in microgravity. Note that a 11-stage cascade is really hard to
observe with a traditional optical instrumentation, since the smallest droplets have a radius of about 0.6
micrometers. Moreover, these droplets have a lifetime about 0.5 milliseconds and they probably coalesce in less
than 100 nanoseconds. The continuum media approach is usually valid. Indeed, when working with two liquids, the
mean free path is clearly smaller than the last droplet, about 600 nanometers. And when working with a liquid
surrounded by air, the air viscosity avoids droplets below 10 micrometers.

Data are not in agreement with our predictions concerning an aniline droplet surrounded by water. The aniline
experiences as much as three partial coalescences, while our predictions state for a total coalescence.
Therefore, the critical Ohnesorge $Oh_{1c}$ is higher than 0.02 in this case. The reason is still unknown and
needs further investigations.

%\addcontentsline{toc}{part}{References}
%\bibliography{CoalescencePartielle}

\end{document}